\documentclass[a4paper,10pt]{article}
\usepackage{listings}
\usepackage[english]{babel}
\usepackage[utf8]{inputenc}
\usepackage[table]{xcolor}
\usepackage{authblk}
\usepackage{graphicx}
\usepackage{hyperref}
\usepackage{multirow}

\hypersetup{
    colorlinks=true,
    linkcolor=blue,
    filecolor=magenta,      
    urlcolor=violet,
    citecolor=blue
}

\graphicspath{ {./images/} }
\usepackage{adjustbox}

\definecolor{codegreen}{rgb}{0,0.6,0}
\definecolor{codegray}{rgb}{0.5,0.5,0.5}
\definecolor{codepurple}{rgb}{0.58,0,0.82}
\definecolor{backcolour}{rgb}{0.95,0.95,0.92}

\lstdefinestyle{mystyle}{
    backgroundcolor=\color{backcolour},   
    commentstyle=\color{codegreen},
    keywordstyle=\color{magenta},
    numberstyle=\tiny\color{codegray},
    stringstyle=\color{codepurple},
    basicstyle=\ttfamily\footnotesize,
    breakatwhitespace=false,         
    breaklines=true,                 
    captionpos=b,                    
    keepspaces=true,                 
    numbersep=5pt,                  
    showspaces=false,                
    showstringspaces=false,
    showtabs=false,                  
    tabsize=2
}

\renewcommand{\thefootnote}{\fnsymbol{footnote}}

\usepackage[nottoc]{tocbibind}

\title{Deep artificial neural network for prediction of atrial fibrillation through the analysis of 12-leads standard ECG}
\author[1]{A. Scagnetto \thanks{Corresponding author arjuna.scagnetto@asugi.sanita.fvg.it}}
\author[2]{G. Barbati}
\author[2]{I. Gandin}
\author[1,5]{C. Cappelletto}
\author[3]{G. Baj}
\author[4]{A. Cazzaniga}
\author[4]{F. Cuturello}
\author[4]{A. Ansuini}
\author[3]{L. Bortolussi}
\author[1]{A. Di Lenarda}

\affil[1]{Cardiovascular Department, Azienda Sanitaria Universitaria Giuliano Isontina (ASUGI), Trieste, Italy}
\affil[2]{Biostatistics Unit, Department of Medicine, Surgery and Health Sciences - University of Trieste, Italy}
\affil[3]{Department of Mathematics and Geosciences - University of Trieste, Italy}
\affil[4]{Area Science Park, Research and Technology Institute, Trieste, Italy}
\affil[5]{Department of Medicine, Karolinska Institutet, Stockholm, Sweden}

\begin{document}

\maketitle
\renewcommand{\thefootnote}{\arabic{footnote}}

\medskip

\section*{\normalsize{Abstract}}
\small{Atrial Fibrillation (AF) is a heart's arrhythmia which, despite often asymptomatic, represents an important risk factor for stroke, therefore being able to predict AF at the electrocardiogram exam would be of great impact on actively targeting patients at high risk. In the present work we use Convolution Neural Networks to analyze ECG and predict Atrial Fibrillation starting from realistic datasets, i.e. considering fewer ECG than other studies and extending the maximal distance between ECG and AF diagnosis. We achieved 75.5\% (0.75) AUC firstly increasing our dataset size by a shifting technique and secondarily using the dilation parameter of the convolution neural network. In addition we find that, contrarily to what is commonly used by clinicians reporting AF at the exam,  the most informative leads for the task of predicting AF are D1 and avR. Similarly, we find that the most important frequencies to check are in the range of 5-20 Hz. Finally, we develop a net able to manage at the same time the electrocardiographic signal together with the electronic health record, showing that integration between different sources of data is a profitable path. In fact, the 2.8\% gain of such net brings us to a 78.3\% (std 0.77) AUC. In future works we will deepen both the integration of sources and the reason why we claim avR is the most informative lead. }
\newpage
\section{Introduction}

Atrial Fibrillation (AF) is the most common supraventricular arrhythmia in the general population, affecting especially the elders. Screening of AF in patients $>$65 years old is recommended by European guidelines and definite diagnosis requires a 12-lead standard ECG. In general, AF is a powerful risk factor for stroke, independently increasing risk 5-fold throughout all ages\cite{null_heart_2012}. Since AF is often asymptomatic and frequently clinically undetected, the stroke risk attributed to AF may be substantially underestimated. Therefore, despite its importance, both patients and treating physicians may be unaware of AF presence\cite{null_heart_2012}. Consequently, the development of tools to predict AF from routine and low cost exam like Electrocardiogram (ECG), is bound to have a strong impact on actively targeting patients at high risk. AF can be described as an irregular foci in the atrium that set up chaotic atrial circuits and irregular rapid contraction of the atrium with loss of consistent atrioventricular synchrony due to decremental conduction at the atrio-ventricular node\cite{goodfellow_classification_2017}. 

Deep Learning (DL) is a subfamily of Machine Learning (ML) specified by all those techniques that use the artificial neuron, the perceptron and its evolution, as a computational basis. DL in the last 6-7 years has been in the spotlight both thanks to the increased computing power to which we have access even at the desktop computer level, and to its abstraction abilities\cite{bizopoulos_deep_2019}. Indeed, DL networks are autonomous, they can work directly in the signal space, that is, they deal by themselves with extracting the features on which to do the classification or regression task. Among all DL networks, the Convolution Neural Networks (CNNs) are particularly know being the most useful type of Artificial Neural Network (ANN) for pattern recognition\cite{lecun_backpropagation_1989} \cite{lecun_gradient-based_1998}.

It has been recently shown \cite{hannun_cardiologist-level_2019} that the performance of an ANN in classifying arrhythmia from ECG can exceed that of cardiologists with average experience. Beside this classification task, ANN have already shown good performances in predicting new on-set AF: Raghunath \cite{raghunath_deep_2021} and Attia \cite{attia_artificial_2019} achieved an Area Under the Curve (AUC) about 83\%. 

In this paper, our main goal is to predict new on-set AF by means of CNN applied to ECG, both by themselves and integrated with data from different sources, such as ECG and Electronic Health Records (EHR). Furthermore we investigate here which lead is more informative related to the task of predicting new on-set AF and, lastly, the effect of filtering on performance.

\section{Materials and Methods}

\subsection{Cohort building}
Our database of ECG waveform counts about 160.000 patients and 650.000 ECG's waveforms, spanning a period of 15 years, from 2005 to 2020. The ECG are recorded by the Mortara\texttrademark  devices ELI230 and ELI250 during routine electrocardiograph exams, performed at the Cardiovascular Department of Azienda Sanitaria Universitaria Giuliano Isontina (ASUGI) in Trieste, Italy. By linking these ECG exams with EHR of the Regional Epidemiological Repository (RER) of the Friuli Venezia Giulia region (FVG), we identified patients with new on-set AF during the observation period. RER is a data warehouse that collects all administrative healthcare database of FVG, such as hospitalizations, demographics, drugs prescriptions, laboratory exams and the cardiological clinical evaluations. We excluded patients with AF before 2005 or implanted with Pace Maker (PM) or Implantable Cardiac Defibrillator (ICD) or Cardiac Resynchronization Therapy (CRT).

Since age is a very important factor in developing stroke AF related events, indeed the percentage of strokes attributable to AF increases steeply from 1.5\% at 50 to 59 years of age to 23.5\% at 80 to 89 years of age\cite{null_heart_2012} at a mean age respectively for men of 66.8 years and 74.6 years for women\cite{null_heart_2012}, in this preliminary study, we decided to focus on patients from 60 to 90 years old only, furthermore by this choice we built a cohort as balanced as possible between the two groups, who will develop AF and who will not, avoiding the use of propensity score matching techniques.

To define the event of interest, we selected only patients with almost two episodes of AF, in order to be more specific on the outcome definition.
We used all the ECG of these patients before the first episode of AF with a maximal distance, between ECG's exam and AF, of 3 years. We choose 3 years to maximize the number of available ECG, but at the same time keep the distance between AF diagnosis and ECG exam as short as possible. 
The sample of ECG without AF was built using all ECG of patients without any episode of AF in the observation period, and which are at a maximum distance of 3 year from the censorship times, administrative censoring date, fixed at 31/12/2020, or at patient's death. The final dataset included 50.593 ECG labeled 0AF, meaning that the corresponding patient will not develop AF, and 11.064 ECG labeled 1AF, meaning that they will develop AF.

\subsection{Data preparation}
We worked with 12-lead 10 seconds long ECG. They are sampled at 1kHz, but we re-sampled at 500Hz, primarily for computational purposes. Each ECG is then a matrix of 5000 x 12, the first dimension is time, each point is equal to two millisecond, and second dimension is the lead, going from 1 to 12 and labeled as standard: D1, D2, D3, avR, avF, avL, v1, v2, v3, v4, v5, v6.

\begin{figure}[h!]
    \centering
    \includegraphics[scale=0.4]{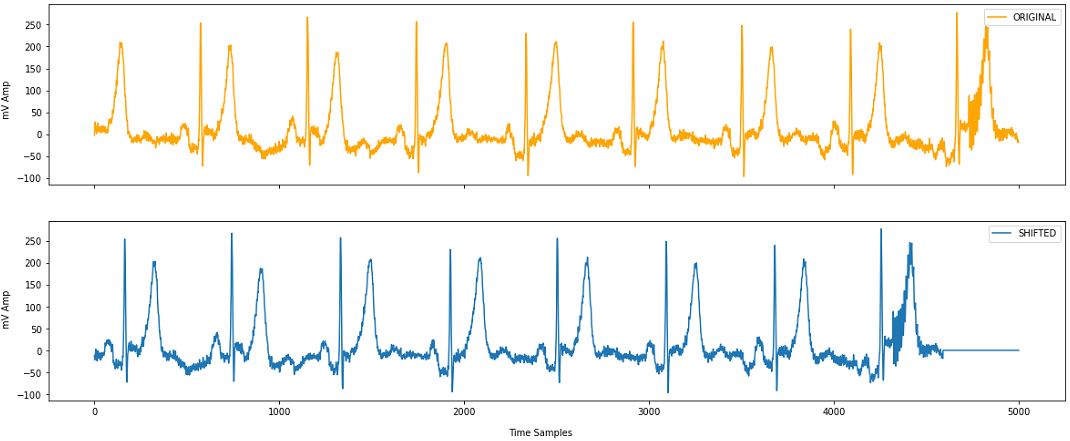}
    \caption{\footnotesize{Original and shifted signal. They both are used in the training set.}}
    \label{fig:shifted_signal}
\end{figure}

We divided the full dataset into training set and test set (test set 12\% of full data-set). Training set was balanced between classes by under sampling 0AF class. The test set, instead, is built following two strategies; a balanced test set used for inner comparability between training runs with different parameters; an unbalanced test set built to mimic the proportion of the classes in the full dataset to get a glimpse of the performance in a real world environment. We used the unbalanced test set only for the general performance measure. Balanced and unbalanced test set contains the same 0AF samples, because we have too few example to differentiate. We augment the number of examples shown to the net in the training phase only, not in testing. Since the CNN are insensitive to rotation and translation, shifting the ECG would enforce the extraction of feature able to distinguish an ECG which leads to AF with respect to the others. Furthermore applying the shifting to each training sample we doubled the number of training examples. The amount of shifting is not fixed, but assigned randomly choosing in a range between 250 and 500 points. Moreover, the shifting is randomly applied at the beginning or at the end of each signal. In figure \ref{fig:shifted_signal} an example of the original and the shifted signal is showed. We do not apply any shifting to the test sets.
In summary, the sets are composed as follows: 
training set is composed of 9737 in class 1FA and their 9737 randomly shifted version, plus 9737 in class 0FA and their randomly 9737 shifted version, summing up to 38948 ECG (19474 1FA and 19474 0FA).
Unbalanced test is composed of 1327 in class 1FA and 6635 in class 0FA, summing up to 7962 ECG.  
Balanced test is composed of 1327 in class 1FA and 1327 in class 0FA, summing up to 2654 ECG. All sets have been shuffled before feeding the net.

To test the \textit{FullModel} net (see Architecture \ref{netshape}) with tabular data too, we used the waves morphology's descriptive parameters, generated by the Mortara\texttrademark devices itself, integrated with patient’s age and gender at the exam. These morphology parameters are: axis, duration, onset, offset of waves P and T and of complex QRS, plus some intervals between waves, PR, QT, RR and finally the cardiac frequency. 

Even if many of these data are strongly correlated, this is not an issue for DL in general e specifically using CNN for classification problems. Indeed we know that DL can scale the dimension of the input into a lower-dimensional space, which hopefully corresponds to the intrinsic dimension\cite{facco_estimating_2017}. 

All tabular data have been normalized and no other cleaning has been performed.

\subsection{Filtering}
The types of noises that afflict the ECG signals are essentially of 4 types: base-line wandering, power-line interference, muscle artifacts, and generic channel noise, such as electrode motion. Each of these noises has its own peculiarities and need specific filters. The most critic noise for AF report is the power-line interference, fortunately Mortara\texttrademark devices have a built-in 50Hz filter, then our ECG are already cleaned. Baseline wander is a low-frequency noise of around 0.5 to 0.6 Hz mainly associated with breathing.

Taking into account that the content of T wave lays mostly within a range from zero (DC) to 10 Hz, P wave is characterized by 5-30 Hz frequencies and QRS usually contains within 8-50 Hz frequencies\cite{tereshchenko_frequency_2015} and even if there are some studies showing that higher frequencies, far beyond 150Hz, contains useful clinical information, we focused on more standard ranges. We tested range 5-50 Hz primarily for comparison with the performance of \textit{EcgNet} with respect to not filtered ECG. Then in order to get some information from the frequencies’s range, we cut the full range in three bands 5-20, 20-35, 35-50, and then we still divided in three the range with the best performance. To apply the filter to the ECG signal we used a fourth order Butterworth band pass. 

\subsection{Architectures}\label{netshape}
We develop two networks, the \textit{EcgNet} working on the ECG only and the \textit{TabNet} for the tabular data only. The \textit{FullModel} merge these two network in one. 

To develop the \textit{EcgNet} we used the net's shape of Goodfellow \cite{goodfellow_towards_2018}. This is a 13-layers net based on the work of Kamaleswaran \cite{kamaleswaran_robust_2018}, but instead of 13 max pooling layers, Goodfellow keeps only 3 layers of max pooling in order to hold the temporal information in the ECG waveform, useful for applying the class activation mapping\cite{zhou_learning_2015}, which is a method to investigate which part of the signal the network is especially interested in. We changed the \textit{dropout} and set them at 5\%, instead of 30\%. 

The \textit{TabNet} is fed with age, gender and the morphology's parameters of the ECG. The net is made of 3 dense layers with 256, 128 and 64 neurons each, and activation is rectified linear unit (RELU). Since dense layer tends to memorize, we added for each dense layer a dropout layer, set to 50\%, and a batch normalization layer. The dimension of last layer, which is 64, has been chosen so to match the dimension of \textit{EcgNet} output in order to add the two outputs. Final the output layer is a 2-neuron dense layer with softmax activation.

\subsection{Training parameters}
We used \textit{cross entropy} loss function, \textit{adam} optimizer, \textit{accuracy} validation metric. To the \textit{learning rate} we applied a decay starting from 0.01 and halving every 15 epochs. We used early stopping to avoid over fitting, configured with patience 15 and minimum delta of 0.005. 

\subsection{Performance metrics}
We used two metrics of performance. Overall accuracy (ACC) that corresponds to the ratio of true positive plus the true negative overall and the Area Under the Curve (AUC), that is the probability that a randomly selected subject with the condition has a test result indicating greater risk than that of a randomly chosen subject without the condition. All performances reported are computed on test sets only.

\subsection{Development environment}
All codes were written in Python 3.5 with TensorFlow 2.4 and others libraries such as: Scipy, Pandas, Numpy, Sklearn. Part of the study has been conducted on a workstation assembled with the new Nvidia RTX 3090 24 Gb, while more massive calculation has been performed on the Genomics platform of Area Science Park with in the "Argo" system.

\section{Results}

\subsection{ECG descriptive statistics}

In table \ref{tab:pop_descriptive} are shown the descriptive statistics of the tabular data, used to feed the \textit{TabNet}. Due to the dataset dimension, all the p-values are lower than $0.001$, even if from the clinical point of view no important difference between the two groups is observed.
For continuous parameters the median and interquartile range are reported, while for the categorical parameter (gender) the absolute number and relative frequency.

\begin{table}[h!]
    \centering
    \begin{tabular}{|l|c|c|}
\hline
 feature & 0FA & 1FA  \\
\hline
n & 50593 & 11064  \\
gender = M (\%) & 26883 (53.1) & 6355 (57.4)   \\
age  & 75 [68, 81] & 76 [70, 81]   \\
P\_AXIS  & 59 [42, 71] & 61 [42, 74]   \\
P\_DUR  & 119 [108, 127] & 118 [100, 130] \\
P\_ONSET  & 283 [259, 302] & 270 [241, 293]   \\
P\_OFFSET  & 401 [376, 419] & 386 [355, 411]  \\
PR\_INT  & 169 [152, 189] & 179 [159, 202]   \\
QRS\_AXIS  & 18 [-17, 54] & 13 [-24, 51] \\
QRS\_DUR  & 98 [89, 111] & 102 [92, 120] \\
QRS\_ONSET  & 453 [446, 458] & 450 [443, 457]   \\
QRS\_OFFSET & 551 [543, 559] & 553 [545, 565]   \\
QT\_INT  & 393 [369, 420] & 406 [381, 433]  \\
QTC\_INT  & 415 [399, 434] & 421 [405, 441]  \\
RR\_INTERVAL  & 831 [727, 944] & 871 [766, 986]    \\
T\_AXIS & 57 [36, 73] & 59 [33, 78]   \\
T\_OFFSET  & 845 [821, 870] & 855 [831, 880]   \\
V\_RATE  & 72 [63, 82] & 68 [60, 78] \\

\hline
\end{tabular}
    \caption{\footnotesize{ECG tabular data statistics comparison between the classes 0AF and 1AF. Axis are expressed in degree, duration in milliseconds, V\_RATE in beats per minute and age in years.}}
    \label{tab:pop_descriptive}
\end{table} 

\subsection{General Performance}

Using \textit{EcgNet} only we achieved a ACC performance of 69 (1.05) and 75.5 (0.75) of AUC on the balanced test set and an ACC of 70.3 (0.25) and 76.2 (0.5) of AUC on the unbalanced test set. Using \textit{TabNet} only we achieved ACC 65.7 (0.78) and AUC 71.6 (0.77). 
Finally integrating the two sources of data in the \textit{Full Model} we achieved 72.3 (0.83) of ACC and 78.3 (0.77) of AUC on balanced test set and 71.3 (0.81) ACC and 77.1 (0.69) AUC on unbalanced test set. 

Standard deviation of these results, reported in brackets, are computed on balanced and unbalanced test set for 5 different trained net.

\subsection{1-Lead and filtering performances}
The performance on the single lead is measured as average of 5 training sessions for each single lead. In table \ref{tab:filtering_performance} we reported the AUC only.
\begin{figure}[h!]
    \centering
    \includegraphics[scale=0.39]{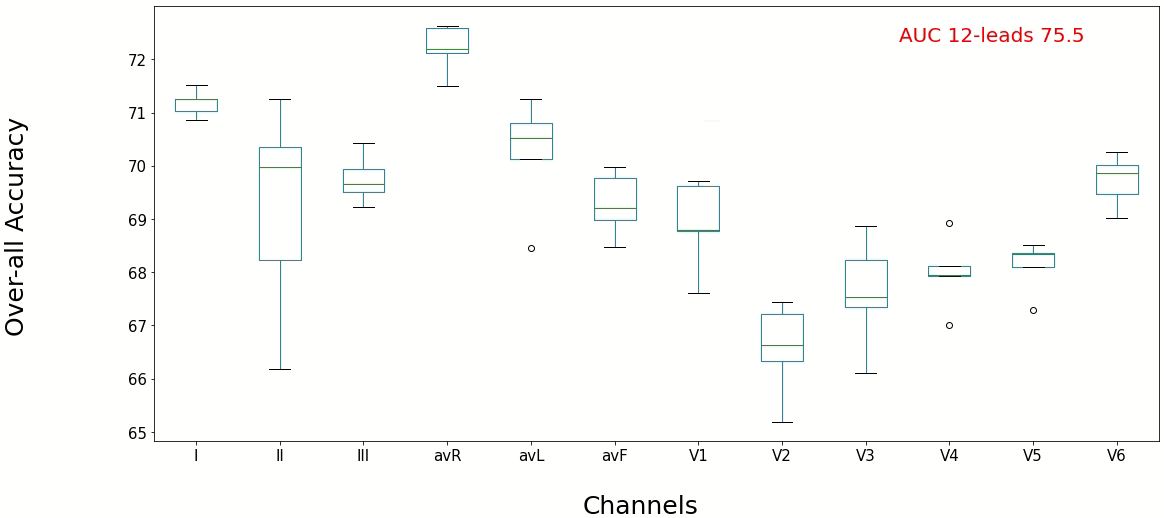}
    \caption{\footnotesize{Box plot of the performances of the \textit{EcgNet} using 1-lead only at a time. In upper right corner in red ink we reported the performance of \textit{EcgNet} on 12-lead input.}}
    \label{fig:lead_performance}
\end{figure}

\begin{table}[h!]
    \centering
    \begin{tabular}{|r||l|l|l|l|l|l|l|l|l|l|l|l|l|l|}
\hline \rowcolor{lightgray}
leads & d1 & d2 & d3 & avR & avF & avL   \\
\hline
Mean & \textcolor{red}{71.2} & 69.2 & 69.7 &\textcolor{red}{72.2} & 70.2 & 69.3 \\
std  & 0.25 & 2.01 & 0.46 & 0.46 & 1.08 & 0.60 \\
\hline \rowcolor{lightgray}
leads & v1 & v2 & v3 & v4 & v5 & v6  \\
\hline
Mean  & 68.9&66.7 &67.6 & 68 & 68.1 & 69.7\\
std  &  0.85 & 0.88 & 1.04 & 0.7 & 0.49 & 0.49\\

\hline
    \end{tabular}
    
    \caption{\footnotesize{Mean and std of AUC on balanced test set averaged on 5 different training. In red the two highest 1-lead performances at D1 and avR.}}
    \label{tab:leads_performance}
\end{table}

In table \ref{tab:filtering_performance} are reported the mean and standard deviation for each band pass range performance averaged on 5 different training session. We reported AUC on balanced test set only. 
\begin{table}[h!]
    \centering
    \begin{tabular}{|r|c|c||c|c|c|}
\hline
 &  \multicolumn{5}{|c|}{band pass range (Hz)} \\\hline
 & unfiltered & [5-50] & [5-20] & [20-35] & [35-50]   \\
\hline
mean & 75.5	& 75.2	& 73.6	& 66.4 & 64.0	\\
std & 0.75 & 0.87	& 0.84	& 1.4 &	0.6\\
\hline
    \end{tabular}
\vspace*{0.75cm}
\begin{tabular}{|r|c|c|c|}
\hline 
& \multicolumn{3}{|c|}{band pass range (Hz)}\\
\hline
 & [5-10] & [10-15] & [15-20]   \\
\hline
mean & 71.6	& 70.0	& 66.0\\
std & 0.5 & 0.7	& 0.2	 	\\
\hline
    \end{tabular}
    \caption{\footnotesize{Mean and std of performance on balanced test set for 5 different training, measured with AUC, using different range of filtering, expressed in Hz.}}
    \label{tab:filtering_performance}
\end{table}
As stated in methods we further cut in three the most informative range of frequencies and we tested the ranges [5-10], [10-15] and [15-20]. The best performance is achieved using the lowest range of frequencies, 5-10 Hz.



\section{Discussion}

\subsection{General Performance}

Our performance if compared with those of Raghunath's \cite{raghunath_deep_2021} and Attia's \cite{attia_artificial_2019} studies, may seems very low, 75.5\% versus 83\%. But firstly they worked at 1 year and 1 month distance between ECG exam and new on-set AF, while we span 3 years. Furthermore they have an enormous dataset about 1.6M ECG and 650K ECG respectively, instead we have only 60 thousand ECG and of which only 11 thousand labeled as new on-set AF. 
We managed this limitations by augmenting the number of ECG, applying a shifting to the ECG waveform and using a net with a particular setting of the dilation parameter. 

Furthermore we investigated the possibility to use other information in order to achieve better results. We used the waves morphology's descriptive, built by the Mortara\texttrademark devices, plus age and gender and using these tabular data together with ECG, we achieved 78.3\% AUC. It's a net gain of 2.8 percentage points. We think that's a promising improvement, that let us wonder that integrating information from different sources represents a very fruitful path to be followed. 





\subsection{1-lead and filtering performances}

In table \ref{tab:leads_performance} we see that D1 and avR have the best performance compared to all others leads. We interpreted this result as an indication that \textit{EcgNet} can gain more information about new on-set AF from those two leads. Since AF is a deformation of the P-wave, which tends to disappear, a logical expectation was that the most of information would have been derived from the morphology characteristic of the P-wave. On the contrary, the median P-wave axis was around 60degree in both 0AF and 1AF groups (in both cases a quite large variability was observed). Looking at the correlation between net class probability assignments and the axis of P and QRS, we noted that for the class assignments with probability greater than 0.90,  either correctly or incorrectly classified,  the QRS complex axis range between 30degree and 10 degree, which are exactly the directions spanned by D1 and -avR . Then we argued that QRS complex axis has a bigger role than the P wave axis in predicting AF.

\begin{table}[h!]
    \centering
\begin{tabular}{|c|c|c|c|}
\hline
& & 0AF & 1AF \\
\hline
\multirow{2}{5em}{Correctly Classified} &P\_AXIS  &      58 [41, 71] &   66 [48, 80]\\
&QRS\_AXIS&     24 [-15, 58] &  20 [-30, 53]\\\hline
\multirow{2}{5em}{Incorrectly classified}&P\_AXIS   &    58 [40, 65]   & 66 [49, 81]\\
&QRS\_AXIS &   24 [-28, 53]  & 15 [-22, 58]\\
\hline
\end{tabular}
    \caption{\footnotesize{Correctly and incorrectly classified median and IQR for axis of P wave and of QRS complex.}}
    \label{tab:probqrs}
\end{table}
The \textit{EcgNet}'s performance on the filtered ECG is 75.2\% AUC (table \ref{tab:filtering_performance}). These proves that for our specific task of predicting new on-set AF, filtering the signal it's not very useful, but in the case filtering is mandatory, such as when it's needed to detect the R peaks, we now know that the range of 5-50 Hz works fine. Moreover we can say that the most useful frequencies are in the lower range around 5 and 20 Hz.

\section{Conclusion and future perspectives}

In literature there's a score useful to predict AF, it's the CHARGE-AF score\cite{alonso_simple_2013}. They built a 5-year predictive model including the variables age, race, height, weight, systolic and diastolic blood pressure, current smoking, use of antihypertensive medication, diabetes, and history of myocardial infarction and heart failure, resulting in a discriminatory level of 76.7 AUC.
It's interesting that at the moment they published the study, we were only at the beginning of the Deep learning era and in that work, authors affirmed that adding PR-interval and ECG-derived left ventricular hypertrophy didn't significantly improve the performance of the score. Instead in the last years, Raghunath\cite{raghunath_deep_2021}, Attia\cite{attia_artificial_2019} has clearly shown that in the ECG there's a lot of information that could be used to predict AF. In future works we plan to test the Full Model using the same tabular data, that Alonso \cite{alonso_simple_2013} included in CHARGE-AF score, with the aim of further improving the overall accuracy. Moreover, we will try to increase both the range of ages and of the distance between ECG exam and incidence of AF, hoping that when the EcgNet can't find anything useful, because the distance to AF is too wide, then the age and the other variables describing the patient's characteristics will support the decision and thus getting closer to building a deep learning model that is able to integrate information from very different sources. 

In this work we have not treated the aspect of the predictions calibration, but from preliminary analyses we estimated that the calibration function could be almost fitted by a straight line, and therefore the transportability to future data should not be a major issue.


From the 1-lead and filtering experiments we got different insights, indeed the 1-single result suggests that the QRS complex is more informative than the P wave, while from the frequencies analysis we get the exact opposite, that the P wave is more important than the QRS complex. In future work we will focus more deeply in investigate which part of the ECG the net focus on in order to make the class assignation. We will use the Class Activation Mapping (CAM) approach, because the \textit{EcgNet} is already ready to be used for CAM, indeed, as explained in Goodfellow \cite{goodfellow_classification_2017}, preserving the temporal dimension in the immediately layer before Global Average Pooling (GAP) is required in order to apply CAM.

In conclusion, thanks to the EcgNet and to sample's augmentation by shifting, we built a net with a reasonable performance in a real-world setting of a moderate size population, and even if we still need to face the fundamental aspect of calibration, we think that it's a promising tool to help cardiologists in predicting new-onset AF and actively targeting patients at high risk.

\section*{Acknowledgements}
We thanks Giovanni Renna from Hillrom Italia for his support with ECG waveforms extraction and ``Amici del Cuore'', for supporting the purchase of the workstation.



\medskip

\bibliographystyle{IEEEtran}
\bibliography{IEEEabrv,text}

\begin{thebibliography}{10}
\providecommand{\url}[1]{#1}
\csname url@samestyle\endcsname
\providecommand{\newblock}{\relax}
\providecommand{\bibinfo}[2]{#2}
\providecommand{\BIBentrySTDinterwordspacing}{\spaceskip=0pt\relax}
\providecommand{\BIBentryALTinterwordstretchfactor}{4}
\providecommand{\BIBentryALTinterwordspacing}{\spaceskip=\fontdimen2\font plus
\BIBentryALTinterwordstretchfactor\fontdimen3\font minus
  \fontdimen4\font\relax}
\providecommand{\BIBforeignlanguage}[2]{{%
\expandafter\ifx\csname l@#1\endcsname\relax
\typeout{** WARNING: IEEEtran.bst: No hyphenation pattern has been}%
\typeout{** loaded for the language `#1'. Using the pattern for}%
\typeout{** the default language instead.}%
\else
\language=\csname l@#1\endcsname
\fi
#2}}
\providecommand{\BIBdecl}{\relax}
\BIBdecl

\bibitem{null_heart_2012}
\BIBentryALTinterwordspacing
V.~L. Roger, A.~S. Go, D.~M. Lloyd-Jones, E.~J. Benjamin, J.~D. Berry, W.~B.
  Borden, D.~M. Bravata, S.~Dai, E.~S. Ford, C.~S. Fox, H.~J. Fullerton,
  C.~Gillespie, S.~M. Hailpern, J.~A. Heit, V.~J. Howard, B.~M. Kissela, S.~J.
  Kittner, D.~T. Lackland, J.~H. Lichtman, L.~D. Lisabeth, D.~M. Makuc, G.~M.
  Marcus, A.~Marelli, D.~B. Matchar, C.~S. Moy, D.~Mozaffarian, M.~E.
  Mussolino, G.~Nichol, N.~P. Paynter, E.~Z. Soliman, P.~D. Sorlie,
  N.~Sotoodehnia, T.~N. Turan, S.~S. Virani, N.~D. Wong, D.~Woo, and M.~B.
  Turner, ``Heart {Disease} and {Stroke} {Statistics}—2012 {Update},''
  \emph{Circulation}, vol. 125, no.~1, pp. e2--e220, Jan. 2012, publisher:
  American Heart Association. [Online]. Available:
  \url{https://www.ahajournals.org/doi/10.1161/cir.0b013e31823ac046}
\BIBentrySTDinterwordspacing

\bibitem{goodfellow_classification_2017}
S.~D. Goodfellow, A.~Goodwin, R.~Greer, P.~C. Laussen, M.~Mazwi, and D.~Eytan,
  ``Classification of atrial fibrillation using multidisciplinary features and
  gradient boosting,'' in \emph{2017 {Computing} in {Cardiology} ({CinC})},
  Sep. 2017, pp. 1--4, iSSN: 2325-887X.

\bibitem{bizopoulos_deep_2019}
P.~Bizopoulos and D.~Koutsouris, ``\BIBforeignlanguage{eng}{Deep {Learning} in
  {Cardiology}},'' \emph{\BIBforeignlanguage{eng}{IEEE reviews in biomedical
  engineering}}, vol.~12, pp. 168--193, 2019.

\bibitem{lecun_backpropagation_1989}
\BIBentryALTinterwordspacing
Y.~LeCun, B.~Boser, J.~S. Denker, D.~Henderson, R.~E. Howard, W.~Hubbard, and
  L.~D. Jackel, ``Backpropagation applied to handwritten zip code
  recognition,'' \emph{Neural Computation}, vol.~1, no.~4, pp. 541--551, Dec.
  1989. [Online]. Available: \url{https://doi.org/10.1162/neco.1989.1.4.541}
\BIBentrySTDinterwordspacing

\bibitem{lecun_gradient-based_1998}
Y.~Lecun, L.~Bottou, Y.~Bengio, and P.~Haffner, ``Gradient-based learning
  applied to document recognition,'' \emph{Proceedings of the IEEE}, vol.~86,
  no.~11, pp. 2278--2324, Nov. 1998, conference Name: Proceedings of the IEEE.

\bibitem{hannun_cardiologist-level_2019}
\BIBentryALTinterwordspacing
A.~Y. Hannun, P.~Rajpurkar, M.~Haghpanahi, G.~H. Tison, C.~Bourn, M.~P.
  Turakhia, and A.~Y. Ng, ``\BIBforeignlanguage{en}{Cardiologist-level
  arrhythmia detection and classification in ambulatory electrocardiograms
  using a deep neural network},'' \emph{\BIBforeignlanguage{en}{Nature
  Medicine}}, vol.~25, no.~1, pp. 65--69, Jan. 2019, number: 1 Publisher:
  Nature Publishing Group. [Online]. Available:
  \url{https://www.nature.com/articles/s41591-018-0268-3}
\BIBentrySTDinterwordspacing

\bibitem{raghunath_deep_2021}
\BIBentryALTinterwordspacing
S.~Raghunath, J.~M. Pfeifer, A.~E. Ulloa-Cerna, A.~Nemani, T.~Carbonati,
  L.~Jing, D.~P. vanMaanen, D.~N. Hartzel, J.~A. Ruhl, B.~F. Lagerman, D.~B.
  Rocha, N.~J. Stoudt, G.~Schneider, K.~W. Johnson, N.~Zimmerman, J.~B. Leader,
  H.~L. Kirchner, C.~J. Griessenauer, A.~Hafez, C.~W. Good, B.~K. Fornwalt, and
  C.~M. Haggerty, ``\BIBforeignlanguage{en}{Deep {Neural} {Networks} {Can}
  {Predict} {New}-{Onset} {Atrial} {Fibrillation} {From} the 12-{Lead}
  {Electrocardiogram} and {Help} {Identify} {Those} at {Risk} of {AF}-{Related}
  {Stroke}},'' \emph{\BIBforeignlanguage{en}{Circulation}}, p.
  CIRCULATIONAHA.120.047829, Feb. 2021. [Online]. Available:
  \url{https://www.ahajournals.org/doi/10.1161/CIRCULATIONAHA.120.047829}
\BIBentrySTDinterwordspacing

\bibitem{attia_artificial_2019}
\BIBentryALTinterwordspacing
Z.~I. Attia, P.~A. Noseworthy, F.~Lopez-Jimenez, S.~J. Asirvatham, A.~J.
  Deshmukh, B.~J. Gersh, R.~E. Carter, X.~Yao, A.~A. Rabinstein, B.~J.
  Erickson, S.~Kapa, and P.~A. Friedman, ``\BIBforeignlanguage{English}{An
  artificial intelligence-enabled {ECG} algorithm for the identification of
  patients with atrial fibrillation during sinus rhythm: a retrospective
  analysis of outcome prediction},'' \emph{\BIBforeignlanguage{English}{The
  Lancet}}, vol. 394, no. 10201, pp. 861--867, Sep. 2019, publisher: Elsevier.
  [Online]. Available:
  \url{https://www.thelancet.com/journals/lancet/article/PIIS0140-6736(19)31721-0/fulltext}
\BIBentrySTDinterwordspacing

\bibitem{facco_estimating_2017}
\BIBentryALTinterwordspacing
E.~Facco, M.~d’Errico, A.~Rodriguez, and A.~Laio,
  ``\BIBforeignlanguage{en}{Estimating the intrinsic dimension of datasets by a
  minimal neighborhood information},'' \emph{\BIBforeignlanguage{en}{Scientific
  Reports}}, vol.~7, no.~1, p. 12140, Sep. 2017, bandiera\_abtest: a
  Cc\_license\_type: cc\_by Cg\_type: Nature Research Journals Number: 1
  Primary\_atype: Research Publisher: Nature Publishing Group Subject\_term:
  Applied mathematics;Computational biophysics;Data mining Subject\_term\_id:
  applied-mathematics;computational-biophysics;data-mining. [Online].
  Available: \url{https://www.nature.com/articles/s41598-017-11873-y}
\BIBentrySTDinterwordspacing

\bibitem{tereshchenko_frequency_2015}
\BIBentryALTinterwordspacing
L.~G. Tereshchenko and M.~E. Josephson, ``Frequency {Content} and
  {Characteristics} of {Ventricular} {Conduction},'' \emph{Journal of
  electrocardiology}, vol.~48, no.~6, pp. 933--937, 2015. [Online]. Available:
  \url{https://www.ncbi.nlm.nih.gov/pmc/articles/PMC4624499/}
\BIBentrySTDinterwordspacing

\bibitem{goodfellow_towards_2018}
\BIBentryALTinterwordspacing
S.~D. Goodfellow, A.~Goodwin, R.~Greer, P.~C. Laussen, M.~Mazwi, and D.~Eytan,
  ``\BIBforeignlanguage{en}{Towards {Understanding} {ECG} {Rhythm}
  {Classification} {Using} {Convolutional} {Neural} {Networks} and {Attention}
  {Mappings}},'' in \emph{\BIBforeignlanguage{en}{Proceedings of the 3rd
  {Machine} {Learning} for {Healthcare} {Conference}}}.\hskip 1em plus 0.5em
  minus 0.4em\relax PMLR, Nov. 2018, pp. 83--101, iSSN: 2640-3498. [Online].
  Available: \url{https://proceedings.mlr.press/v85/goodfellow18a.html}
\BIBentrySTDinterwordspacing

\bibitem{kamaleswaran_robust_2018}
R.~Kamaleswaran, R.~Mahajan, and O.~Akbilgic, ``\BIBforeignlanguage{eng}{A
  robust deep convolutional neural network for the classification of abnormal
  cardiac rhythm using single lead electrocardiograms of variable length},''
  \emph{\BIBforeignlanguage{eng}{Physiological Measurement}}, vol.~39, no.~3,
  p. 035006, Mar. 2018.

\bibitem{zhou_learning_2015}
\BIBentryALTinterwordspacing
B.~Zhou, A.~Khosla, A.~Lapedriza, A.~Oliva, and A.~Torralba, ``Learning {Deep}
  {Features} for {Discriminative} {Localization},'' \emph{arXiv:1512.04150
  [cs]}, Dec. 2015, arXiv: 1512.04150. [Online]. Available:
  \url{http://arxiv.org/abs/1512.04150}
\BIBentrySTDinterwordspacing

\bibitem{alonso_simple_2013}
A.~Alonso, B.~P. Krijthe, T.~Aspelund, K.~A. Stepas, M.~J. Pencina, C.~B.
  Moser, M.~F. Sinner, N.~Sotoodehnia, J.~D. Fontes, A.~C. J.~W. Janssens,
  R.~A. Kronmal, J.~W. Magnani, J.~C. Witteman, A.~M. Chamberlain, S.~A.
  Lubitz, R.~B. Schnabel, S.~K. Agarwal, D.~D. McManus, P.~T. Ellinor, M.~G.
  Larson, G.~L. Burke, L.~J. Launer, A.~Hofman, D.~Levy, J.~S. Gottdiener,
  S.~Kääb, D.~Couper, T.~B. Harris, E.~Z. Soliman, B.~H.~C. Stricker,
  V.~Gudnason, S.~R. Heckbert, and E.~J. Benjamin,
  ``\BIBforeignlanguage{eng}{Simple risk model predicts incidence of atrial
  fibrillation in a racially and geographically diverse population: the
  {CHARGE}-{AF} consortium},'' \emph{\BIBforeignlanguage{eng}{Journal of the
  American Heart Association}}, vol.~2, no.~2, p. e000102, Mar. 2013.

\end{thebibliography}

\end{document}